\documentstyle[12pt,dina4p]{article}

\newcommand{\be}{\begin{equation}}
\newcommand{\ee}{\end{equation}}
\newcommand{\ba}{\begin{eqnarray}}
\newcommand{\ea}{\end{eqnarray}}
\newcommand{\bc}{\begin{center}}
\newcommand{\ec}{\end{center}}

\def\sign{\mbox{\rm sign}}
\def\Tr{\mbox{\rm Tr}}
\def\CS{\mbox{\rm \scriptsize  CS}}
\def\ren{\mbox{\rm \scriptsize  ren}}
\def\bare{\mbox{\rm \scriptsize bare}}
\def\D{\not\!\!D}
\def\DD{\not\!\!\!D}

\newcommand{\aq}{\bar{A}}
\newcommand{\gk}{$\Gamma_k$}
\newcommand{\mez}{\frac{1}{2}}
\newcommand{\NA}{\mbox{\scriptsize {\rm NA}}}
\newcommand{\conf}{\mbox{\scriptsize {\rm conf}}}
\begin{document}
\begin{flushright}
hep-th/9602012   \\
DESY 96-016
\end{flushright}   \vspace{10mm}
\begin{center}
{{\Large \bf
Effective Average Actions and       \\
Nonperturbative Evolution Equations
\footnote{Talk given at the 5{\it th} Hellenic School and Workshops
          on Elementary Particle Theory, Corfu, Greece, 1995,
          to appear in the proceedings.}
}}\\
\vspace{5mm}
{\sc M.\, Reuter}\\
{\it Deutsches Elektronen-Synchrotron DESY,\\
     Notkestrasse 85, D-22603 Hamburg, Germany}
\end{center}
\vspace{1mm}
\begin{abstract}
The effective average actions for gauge theories and the associated
nonperturbative evolution equations which govern their renormalization
group flow are reviewed and various applications are described.
As an example of a topological field theory,
Chern-Simons theory is discussed in detail.
\end{abstract}

\renewcommand{\theequation}{1.\arabic{equation}}
\setcounter{equation}{0}
\section*{1 Introduction}

In these notes we first give
a brief introduction to the method of the effective
average actions and their associated exact renormalization
group or evolution equations \cite{wil, quattro, ex},
and then we illustrate these ideas
by means of two examples.
We discuss the renormalization group behavior of the nonabelian
gauge coupling in ``ordinary" Yang-Mills theories and of the
Chern-Simons parameter in pure 3-dimensional Chern-Simons
theory, which provides a first example of a topological field theory.

The effective average action
\gk~can be thought of as a continuum version of the
block spin action for spin systems \cite{wil}.
The functional $\Gamma_k$~is the action relevant to the physics
at (mass) scale $k$. It has the quantum fluctuations with
momenta larger than $k$~integrated out already, but
those with momenta smaller than $k$~are not yet included.
\gk~interpolates between the classical action $S$~
for large values of $k$, and the conventional effective action
for $k$~approaching zero:
$\Gamma_{k\rightarrow\infty} =S ,
 \Gamma_{k\rightarrow 0} = \Gamma$.
     In many important cases
where perturbation theory is inapplicable due to infrared divergences
the limit
$k\rightarrow 0$~exists and can be computed by various methods.
This includes for instance massless theories in low dimensions
or the high temperature limit of 4 dimensional theories.
The functional \gk~can be obtained by solving an exact renormalization
group equation which describes its evolution while $k$~is lowered from
infinity to zero. In the approach of ref. \cite{quattro}, and
for models with a scalar field $\phi$ only, this evolution
equation reads
\be\label{S.1}
\frac{\partial}{\partial t}\Gamma_k[\phi]=\frac{1}{2}{\rm Tr}
\left[\frac{\partial}{\partial t} R_k\ \left(\Gamma^{(2)}_k[\phi]
+R_k\right)^{-1}\right]
\ee
Here $t\equiv\ln k$ is the ``renormalization group time'' and
$\Gamma^{(2)}_k$
denotes the matrix of the second functional derivatives of \gk.
The operator $R_k\equiv R_k(-\partial^2)$ or, in momentum space,
$R_k\equiv R_k(q^2)$~describes the details of how the small momentum
modes are cut off and it is to some extent arbitrary. It has to
vanish for $q^2\gg k^2$~and to become a mass-like term
proportional to $k^2$~for small momenta $q^2\ll k^2$.
The derivation of (\ref{S.1}) proceeds as follows.
In the euclidean functional integral for the generating functional
of the connected Green functions one adds a momentum-dependent
mass term (playing the role of a smooth IR cutoff)
$\mez \int \phi R_k(-\partial^2)\phi$
to the classical action $S$. Then, up to an explicitly known
correction term \cite{quattro}, the resulting $k$-dependent functional
$W_k[J]$
is related to
$\Gamma_k[\phi]$
by a conventional Legendre transformation at fixed $k$.

In Section 2 we generalize the above evolution equation
to gauge theories, and in Section 3 we discuss its BRS properties.
In Section 4 we show how it can be used to
calculate the beta-function of the nonabelian gauge
coupling. In Sections 5 and 6 we shall apply the same
strategies to the study of pure
Chern-Simons field theory in 3 dimensions.

\renewcommand{\theequation}{2.\arabic{equation}}
\setcounter{equation}{0}
\section*{2 The Renormalization Group Equation}

In the case of gauge theories the derivation of an exact evolution
equation faces additional complications because the inhomogeneous
gauge transformation law of the Yang-Mills fields
forbids a mass-type cutoff. In refs.
\cite{ex,ahm} this problem was overcome recently
by using the background gauge technique \cite{abb}
which allows us to work with a {\it gauge invariant}
effective average action. The price which one has to pay for this
advantage is that
$\Gamma_k$ depends on two gauge fields: the usual
classical average field $A^a_\mu$ and the background field
$\bar{A}^a_\mu$.
For pure Yang-Mills theory one finds the following
renormalization group equation \cite{ex}
\ba        \label{3}
     k\frac{d}{dk} \Gamma_k[A,\bar{A}] & = & \frac{1}{2} \Tr
\left[\left(\Gamma_k^{(2)}[A,\bar{A}] +R_k(\Delta[\bar{A}])\right)^{-1}
     k \frac{d}{dk}R_k(\Delta[\bar{A}])\right]
               \\
     && -\Tr\left[\left(-D_{\mu}[A]\, D_{\mu}[\bar{A}] +
     R_k(-D^2(\bar{A}))\right)^{-1} k \frac{d}{dk}
     R_k(-D^2[\bar{A}])\right]    \nonumber
\ea
In writing down this equation we made a certain approximation
on which we shall be more explicit in Section 3 where we also sketch
the details of its derivation.
Eq.(\ref{3}) has to be solved subject to the initial condition
\be  \label{4}
     \Gamma_{\infty}[A,\bar{A}] =S[A]+\frac{1}{2\alpha} \int
     d^d\!x~
 \left(D^{ab}_{\mu}[\bar{A}]~(A_{\mu}^b-\bar{A}_{\mu}^b)\right)^2
\ee
where the classical action is augmented by the background gauge
fixing term. Furthermore, $\Gamma^{(2)}_k[A,\bar{A}]$ denotes the
matrix of the second functional derivatives of $\Gamma_k$ with
respect to $A$ at fixed $\bar{A}$.
Again, the function $R_k$ specifies the precise form of
the infrared cutoff, and it has the same properties as mentioned
in the introduction. A convenient choice is
\be   \label{5}
     R_k(u)=Z_k\ u~\left[\exp{(u/k^2)}-1\right]^{-1}
\ee
but in some cases even a simple constant $R_k=Z_k k^2$ is sufficient.
The factor $Z_k$~has to be fixed in such a way that a massless
inverse propagator $Z_k q^2$~combines with the cutoff to
$Z_k(q^2+k^2)$~for the low momentum modes.
$Z_k$~may be chosen differently for different fields.
In particular, different $Z_k$-factors are used for the gauge field
fluctuations and for the Faddeev-Popov ghosts. (They give rise to
the first and the second trace on the RHS of eq.(\ref{3}),
respectively.)
Observable quantities will not depend on the form of $R_k$. A
similar remark applies to the precise form of the operator
$\Delta[\bar{A}] \equiv -D^2 [\bar{A}]+...$ which is essentially
the covariant laplacian, possibly with additional nonminimal
terms \cite{ex}. The r\^{o}le of $\Delta$ is to distinguish
``high momentum modes" from ``low momentum modes". If one expands
all quantum fluctuations in terms of the eigenmodes of $\Delta$,
then it is the modes with eigenvalues larger than $k^2$ which are
integrated out in $\Gamma_k$.

In order to understand the structure of the renormalization
group equation (\ref{3}) it is useful to realize that it can be
rewritten in a form which is reminiscent of a one-loop formula:
\ba\label{1.3}
\frac{\partial}{\partial t} \Gamma_k[A,\bar
A]&=&\frac{1}{2}\frac{D}{Dt}\Tr\ln\left[\Gamma^{(2)}_k[A,\bar
A]+R_k\left(\Delta[\bar A]\right)\right]\nonumber\\
&&-\frac{D}{Dt}\Tr\ln\left[-D^\mu[A]D_\mu[\bar A]+R_k(-D^2[\bar
A])\right]
\ea
By definition, the derivative
$\frac{D}{Dt}$ acts only on the explicit $k$-dependence of
$R_k$, but not on $\Gamma^{(2)}_k [A,\bar A]$.  It is easy now
to describe the relation between the effective average action
$\Gamma_k$ and the conventional effective action. Let us first make the
approximation $\frac{D}{Dt}\to\frac{\partial}{\partial t}$ in eq.
(\ref{1.3}). This amounts to neglecting the running of $\Gamma_k$ on the
RHS of the evolution equation. Therefore it can be solved
by simply integrating both sides of the equation from the
infrared cutoff $k$~to the ultraviolet cutoff $\Lambda$:
\ba\label{1.4}
\Gamma_k[A,\bar A]&=&\Gamma_\Lambda[A,\bar
A]+\frac{1}{2}\Tr\left\{\ln\left[\Gamma_k^{(2)}[A,\bar
A]+R_k(\Delta[\bar A])\right]\right.\nonumber\\
&&\left.-\ln\left[\Gamma_\Lambda^{(2)}[A,\bar A]+R_\Lambda
(\Delta[\bar A])\right]\right\}\nonumber\\
&&-\Tr\left\{\ln\left[-D^\mu[A]D_\mu[\bar A]+R_k(-D^2[\bar
A])\right]\right.\nonumber\\
&&\left.-\ln\left[-D^\mu[A]D_\mu[\bar A]+R_\Lambda(-D^2[\bar
A])\right]\right\}
+O\left(\frac{\partial}{\partial t}\Gamma^{(2)}_k\right)
\ea
Ultimately we shall send the ultraviolet cutoff to infinity and identify
$\Gamma_\Lambda$ with the classical action $S$ plus the gauge fixing
term. Eq.(\ref{1.4})  has  a similar structure as a
regularized version of the conventional one-loop effective action in
the background gauge. There are two important differences,
however: (i) The second variation of the classical action, $S^{(2)}$, is
replaced by $\Gamma^{(2)}_k$.
This implements a kind of ``renormalization
group improvement''.
(ii)  The effective average action contains an explicit infrared cutoff
$R_k$. Because
\be
\lim_{u\to\infty} R_k(u)=0, \ \ \  \lim_{u\to 0} R_k(u)=Z_k k^2
\ee
a mass-term is added to the inverse propagator
$\Gamma^{(2)}_k$ for low frequency modes $(u\to 0)$, but not
for high frequency modes $(u\to \infty)$.

The solution $\Gamma_k[A,\bar{A}]$ of
(\ref{3})
with (\ref{4}) is gauge invariant under simultaneous gauge
transformations of $A$ and $\bar{A}$.
Following the lines of the conventional background method \cite{abb}
one would try to equate the two gauge field arguments of
\gk, and work with the functional
$\bar\Gamma_k[A]\equiv\Gamma_k[A,A]$.
However, it is important to note that the evolution equation (\ref{%
3}) cannot be rewritten in terms of
$\bar\Gamma_k[A]$
alone, since
$\Gamma_k^{(2)}$
does not involve derivatives with respect to $\aq$. In fact, let us
introduce the decomposition
\be\label{6.9}
\Gamma_k[A,\bar A]
    =\bar\Gamma_k[A]+\Gamma_k^{\rm gauge}[%
A,\bar A]    \ee
This leads to
\be\label{6.11}
\Gamma_k^{(2)}[A,\bar A]%
    =\bar\Gamma_k^{(2)}[A]+\Gamma_k^{{\rm
gauge}(2)}
[A,\bar A]_{|\bar A}\ee
where the second functional derivative
$\Gamma_k^{{\rm
gauge}(2)}$
is performed at fixed $\aq$. The interpretation of (\ref{6.9})
and (\ref{6.11}) is as follows. Because
$\bar\Gamma_k[A]$
is a gauge invariant functional of its argument,
$\bar\Gamma_k^{(2)}[A]$ is necessarily singular, i.e., it has the
usual gauge zero modes. They are gauge fixed by the generalized
gauge fixing term
$\Gamma_k^{\rm gauge}$. This is possible because
$\Gamma_k^{{\rm
gauge}}$ is not invariant under separate gauge transformations
of $A$~alone.
\renewcommand{\theequation}{3.\arabic{equation}}
\setcounter{equation}{0}
\section*{3 BRS-Symmetry and Modified Slavnov-Taylor Identities}

In order to actually derive the evolution equation as well as the
pertinent Ward-Takahashi or
Slavnov-Taylor identities we start
from the following scale dependent generating functional
in the background formalism \cite{abb}:
\ba\label{AA.1}
\exp W_k[K^a_\mu,\sigma^a,\bar\sigma^a;\bar\beta^a_\mu,\bar\gamma^a;\bar
A^a_\mu]&=&\int {\cal D} {\cal A} {\cal D} C{\cal D}\bar C\ \exp-\left\{
S[{\cal A}]+\Delta_kS\right.\nonumber\\
\left. +S_{\rm gf}+S_{\rm ghost}+S_{\rm source}\right\}&\equiv%
       &\int {\cal D}\phi\ \exp(-S_{\rm tot})         \ea
Here $S[{\cal A}]$ denotes the gauge invariant classical action, and
\ba\label{AA.2}
\Delta_kS&=&\frac{1}{2}\int d^dx\ ({\cal A}-\bar A)^a_\mu R_k(\bar
A)^{ab}_{\mu\nu}({\cal A}-\bar A)^b_\nu\nonumber\\
&&+\int d^d x\ \bar C^a R_k(\bar A)^{ab} C^b\ea
is the infrared cutoff (``momentum dependent mass term")
                       for the gauge field fluctuation $a\equiv {\cal A}-\bar
A$ and for the Faddeev-Popov ghosts $C$ and $\bar C$. Here $R_k(\bar A)$ is a
suitable cutoff operator which depends on $\bar A$ only. It may be chosen
differently for the gauge field and for the ghosts.
Furthermore
\be\label{AA.3}
S_{\rm gf}=\frac{1}{2\alpha}%
                        \int d^d x\left[ D_\mu(\bar A)^{ab}({\cal A}-\bar
A)^b_\mu\right]^2\ee
is the background gauge fixing term and
\be\label{AA.4}
S_{\rm ghost}=-\int d^dx \ \bar C^a\left(D_\mu(\bar A) D_\mu({\cal
A})\right)^{ab} C^b\ee
is the corresponding ghost action \cite{abb}.
The fields ${\cal A}-\bar A$, $\bar C$ and $C$ are
coupled to the sources $K,\ \sigma$ and $\bar\sigma$, respectively:
\ba\label{AA.5}
S_{\rm source}&=&-\int d^d x\left\{ K^a_\mu({\cal A}^a_\mu-\bar A_\mu^a)+
\bar\sigma^a C^a+\sigma^a\bar C^a\right.\nonumber\\
&&+\left.\frac{1}{g}\bar\beta^a_\mu D_\mu({\cal A})^{ab} C^b+\frac{1}{2}
\bar\gamma^a f^{abc} C^b C^c\right\}.\ea
We also included the sources $\bar\beta$ and $\bar\gamma$
which couple to the
BRS-variations of ${\cal A}$ and of $C$, respectively. In fact,
$S+S_{\rm gf}+S_{\rm ghost}$ is invariant under the BRS transformation
\ba\label{AA.6}
\delta {\cal A}^a_\mu&=&\frac{1}{g}\varepsilon D_\mu({\cal A})^{ab}
 C^b\nonumber\\
\delta C^a&=&-\frac{1}{2}\varepsilon f^{abc}C^b C^c\nonumber\\
\delta \bar C^a&=&\frac{\varepsilon}{\alpha g} D_\mu (\bar A)^{ab}
({\cal A}_\mu^b-\bar A_\mu^b) \ea
Let us introduce the classical fields
\be\label{AA.7}
\bar a^b_\mu=\frac{\delta W_k}{\delta K_\mu^b},\
\xi^b=\frac{\delta W_k}{\delta \bar\sigma^b},
\ \bar \xi^b=\frac{\delta W_k}{\delta\sigma^b}\ee
and let us formally solve the relations
$\bar a=\bar a(K,\sigma,\bar\sigma;\bar\beta,\bar\gamma;\bar A),
\ \xi=\xi(...)$, etc., for the sources $K,\sigma$ and
$\bar\sigma:\ K=K(\bar a,\xi,\bar \xi;\bar\beta,\bar\gamma;\bar A),\
\sigma=\sigma(...),\ ...$. We introduce the
new functional $\tilde\Gamma_k$ as the Legendre transform of $W_k$ with
respect to $K,\sigma$ and $\bar\sigma$:
\ba\label{AA.8}
\tilde\Gamma_k[\bar a, \xi,\bar \xi;\bar\beta,\bar\gamma;\bar A]&=&
\int d^dx\{K^b_\mu\bar a^b_\mu+\bar\sigma^b \xi^b+
\sigma^b\bar\xi^b\}\nonumber\\
&&-W_k[K,\sigma,\bar\sigma;\bar\beta,\bar\gamma;\bar A].\ea
Apart from the usual relations
\be\label{AA.9}
\frac{\delta\tilde\Gamma_k}{\delta\bar a^a_\mu}=K^a_\mu,\qquad
\frac{\delta\tilde\Gamma_k}{\delta \xi^a}=-\bar\sigma^a,\qquad
\frac{\delta\tilde\Gamma_k}{\delta\bar \xi^a}=-\sigma^a\ee
we have also
\be\label{AA.10}
\frac{\delta\tilde\Gamma_k}{\delta\bar\beta^a_\mu}=-
\frac{\delta W_k}{\delta\bar\beta^a_\mu},\qquad
\frac{\delta\tilde\Gamma_k}{\delta\bar\gamma^a}=-
\frac{\delta W_k}{\delta\bar\gamma^a}\ee
where $\delta\tilde \Gamma / \delta\bar{\beta}$ is
taken for fixed $\bar a,\xi,\bar \xi$ and
$\delta W   /  \delta\bar{\beta}$ for fixed
$K,\sigma,\bar\sigma$. The effective average action $\Gamma_k$ is
obtained by subtracting the IR cutoff $\Delta_k S$,
expressed in terms of the classical fields,
from the Legendre transform $\tilde\Gamma_k$:
\ba\label{AA.11}
\Gamma_k[\bar a,\xi,\bar \xi;\bar\beta,\bar\gamma;\bar A]&=&
\tilde\Gamma_k[\bar a, \xi,\bar \xi;\bar\beta,\bar\gamma;\bar A]
-\frac{1}{2}\int d^dx\ \bar a^a_\mu R_k(\bar A)^{ab}_{\mu\nu}\bar
 a^b_\nu\nonumber\\
&&-\int d^dx\ \bar \xi^a R_k(\bar A)^{ab}\xi^b.\ea
Frequently we shall use the field $A\equiv\bar A+\bar a$ (the classical
counterpart of ${\cal A}\equiv \bar A+a)$ and write correspondingly
\be\label{AA.12}
\Gamma_k[A,\bar A,\xi,\bar \xi;\bar\beta,\bar\gamma]\equiv\Gamma_k[A-
\bar A,\xi,\bar \xi;\bar\beta,\bar\gamma;\bar A].\ee
For $\xi=\bar \xi=\bar\beta=\bar\gamma=0$
one recovers the effective average action
$\Gamma_k[A,\bar A]$ which we discussed in Section 2.

Upon taking the $k$-derivative of eq. (\ref{AA.1}) and Legendre-%
transforming the result
                                  one finds the following exact evolution
equation ($t=\ln k$):
\ba               \label{AA.13}
&&\frac{\partial}{\partial t}\Gamma_k[A,\bar A,\xi,\bar
\xi;\bar\beta,\bar\gamma]=\frac{1}{2}\Tr
\left[\left(\Gamma^{(2)}_k+R_k(\bar A)\right)
^{-1}_{AA}\frac{\partial}{\partial t}R_k(\bar A)_{AA}\right]
\nonumber\\
&&-\frac{1}{2}\Tr\left[\left(\left(\Gamma^{(2)}_k+R_k(\bar
A))\right)^{-1}_{\bar \xi \xi}-\left(\Gamma_k^{(2)}+R_k
(\bar A)\right)^{-1}_{\xi\bar\xi}\right)
\frac{\partial}{\partial t} R_k(\bar A)_{\bar \xi \xi}\right]        \ea
Here $\Gamma^{(2)}_k$ is the Hessian of $\Gamma_k$
with respect to $A,\xi$ and $\bar \xi$ at fixed $\bar A,
\bar\beta$ and $\bar\gamma$ and, in an obvious notation,
                                $R_{kAA},R_{k\bar \xi\xi}$
are the infrared cutoff operators introduced in (\ref{AA.2}).
The evolution equation (\ref{AA.13}) is exact in the sense that
its solution, when evaluated at $k=0$, equals the
{\it exact} generating functional of the 1PI Green's functions in the
background gauge, i.e., it is not just an improved one-loop functional.

It is clear from its construction that $\Gamma_k$ is invariant under
simultaneous gauge transformations of $A_\mu$ and
$\bar A_\mu$ and homogeneous transformations of $\xi,\bar
\xi, \bar\beta_\mu$ and $\bar\gamma$, i.e., $\delta
\Gamma_k[A,\bar A,\xi,\bar \xi;\bar\beta,\bar\gamma]=0$ for
\ba\label{AA.14}
\delta A^a_\mu&=&-\frac{1}{g}D_\mu(A)^{ab} \omega^b\nonumber\\
\delta\bar A^a_\mu&=&-\frac{1}{g} D_\mu(\bar A)^{ab} \omega^b\nonumber\\
\delta V^a&=&f^{abc} V^b \omega^c,\ V\equiv \xi,\bar \xi,
\bar\beta_\mu,\bar\gamma.\ea

Next we turn to the Ward identities. By applying the transformations
(\ref{AA.6}) to the integrand of (\ref{AA.1}) one obtains from the BRS
invariance of the measure ${\cal D}\phi$
\be\label{AA.15}
\int{\cal D}\phi\ \delta_{\rm BRS}\exp(-S_{\rm tot})=0\ee
or
\ba\label{AA.15a}
&&\int d^dx\left\{ K^a_\mu\frac{\delta W_k}{\delta\bar\beta^a_\mu}
+\bar\sigma^a
\frac{\delta W_k}{\delta \bar\gamma^a}-\frac{1}
{\alpha g}\sigma^a D_\mu(\bar A)^{ab}
\frac{\delta W_k}{\delta K^b_\mu}\right\}\nonumber\\
&=&\int d^dx\left\{\left[\frac{\delta
W_k}{\delta\bar\beta^a_\mu}+\frac{\delta}{\delta\bar\beta^a_\mu}
\right]\left( R_k\frac{\delta W_k}{\delta K}\right)^a_\mu+
\frac{1}{\alpha g}\left( D_\mu(\bar A)\left[\frac{\delta W_k}{\delta
K_\mu}+\frac{\delta}{\delta K_\mu}\right]\right)^a\right.
\left(R_k\frac{\delta W_k}{\delta\bar\sigma}\right)^a\nonumber\\
&&\left.+\left[\frac{\delta W_k}{\delta\sigma^a}+
\frac{\delta}{\delta\sigma^a}
\right]\left(R_k\frac{\delta W_k}{\delta\bar\gamma}\right)^a\right\}\ea
with $(R_k\delta W_k/\delta\bar\gamma)^a\equiv R_k
(\bar A)^{ab}\delta W_k/\delta \bar\gamma^b$, etc.
Equation (\ref{AA.15a}) can be converted
to the following relation for the effective average action (\ref{AA.12}):
\be\label{AA.16}
\int d^dx\left\{
\frac{\delta\Gamma'_k}{\delta
 A^a_\mu}\frac{\delta\Gamma'_k}{\delta\bar\beta^a_\mu}
-\frac{\delta\Gamma'_k}{\delta \xi^a}\frac{\delta\Gamma'_k}
{\delta\bar\gamma^a}
\right\}=\Delta^{\rm (BRS)}_k\ee
where the ``anomalous contribution'' $\Delta^{\rm (BRS)}_k$ is given by
\ba\label{AA.16a}
\Delta_k^{\rm (BRS)}&=&\Tr\left[R_k(\bar A)_{A_\mu
A_\nu}(\Gamma^{(2)}_k+R_k)^{-1}_{A_\mu\varphi}
\frac{\delta^2\Gamma'_k}{\delta \varphi \delta
\bar\beta_\nu}\right]\nonumber\\
&&-\Tr\left[ R_k(\bar A)_{\bar \xi
\xi}\left(\Gamma^{(2)}_k+R_k\right)^{-1}_{
\xi\varphi}\frac{\delta^2\Gamma'_k}{\delta\varphi\delta\bar\gamma}
\right]\nonumber\\
&&-\frac{1}{\alpha g}\Tr \left[ D_\mu(\bar
A)\left(\Gamma^{(2)}_k+R_k\right)^{-1}_{A_\mu \bar\xi} R_k(\bar A)_{\bar \xi
\xi}\right]
\ea
and
\be\label{AA.17}
\Gamma'_k\equiv \Gamma_k-\frac{1}{2\alpha}\int d^d x\left[ D_\mu (\bar A)
(A_\mu-\bar A_\mu)\right]^2.\ee
Here $\varphi\equiv (A_\mu,\xi,\bar \xi)$ is summed over on the RHS of
(\ref{AA.16a}).
In deriving eq. (\ref{AA.16a}) we used
\be\label{AA.18}
\left[\frac{\delta}{\delta\bar \xi^a}-g D_\mu(\bar
A)^{ab}\frac{\delta}{\delta\bar\beta^b_\mu}\right]\Gamma_k[A,\bar A,\xi,\bar
\xi;\bar\beta,\bar\gamma]=0\ee
which follows from the equation of motion of the antighost.
Equation (\ref{AA.16}) is the generating relation for the modified Ward
identities which we wanted to derive. In conventional Yang-Mills theory,
without IR-cutoff, the RHS of (\ref{AA.16}) is zero. The traces on the RHS of
(\ref{AA.16}) lead to a violation of the usual Ward identities for
nonvanishing values of $k$. As $k$ approaches zero, $R_k$ and hence
$\Delta^{\rm (BRS)}_k$
vanishes and we recover the conventional Ward-Takahashi
identities
\cite{Ell2}.

The modified Ward identities (\ref{AA.16}) are not the only conditions
which the average action $\Gamma_k$ has to satisfy. There exists also
an exact formula for its
$\bar A$-derivative:
\ba\label{AA.20}
&&\frac{\delta}{\delta\bar A^a_\mu(y)}\Gamma_k'[A,\bar A,\xi,\bar
\xi;\bar\beta,\bar\gamma]
=-g^2\bar \xi^b(y)f^{abc}\frac{\delta\Gamma_k}{\delta\bar\beta^c_\mu(y)}
\nonumber\\
\ \ &&+\frac{1}{2}\Tr\left[\left(\Gamma_k^{(2)}+R_k\right)^{-1}
_{AA}\frac{\delta}{\delta\bar A_\mu^a(y)}\left(R_k-\frac{1}{\alpha}
\bar D\otimes\bar D\right)_{AA}\right]\nonumber\\
\ \ &&-\Tr\left[\left(\Gamma_k^{(2)}+R_k\right)^{-1}_{\bar\xi\xi}
\frac{\delta R_{k\bar\xi\xi}}{\delta\bar A^a_\mu(y)}\right]\nonumber\\
\ \  &&+g^2\int d^dx\ {\rm tr}%
                 \left[T^a \left(\Gamma_k^{(2)}+R_k\right)^{-1}
_{\bar\xi(y)\varphi(x)}\frac{\delta^2\Gamma_k}{\delta\varphi(x)
\delta\bar\beta_\mu(y)}\right]\ea
Note that the RHS of eq. (\ref{AA.20}) does not
vanish even for $k\to0$. The $\bar D\otimes\bar D$-piece of the
2nd term and the 4th term on the r.h.s. of (\ref{AA.20}) survive
this limit.

So far we were deriving general identities which constrain the
form of the exact functional $\Gamma_k$. Let us now
ask what they imply if we truncate the space of actions. It is often
sufficient \cite{ahm,lit}
to neglect the $k$-evolution of the ghost sector by
making an ansatz which keeps the classical form of the corresponding
terms in the action
\be\label{A.23}
\Gamma_k[A,\bar A,\xi, \bar\xi;\bar\beta,\bar\gamma]
=\Gamma_k[A,\bar A]+\Gamma_{\rm gh}\ee
\ba\label{A.24}
\Gamma_{\rm gh}&=&-\int d^dx\ \bar\xi D_\mu(\bar A)D_\mu(A)\xi\nonumber\\
&&-\int d^dx\left\{\frac{1}{g}\bar\beta^a_\mu D_\mu(A)^{ab}
\xi^b+\frac{1}{2}\bar\gamma^a f^{abc}\xi^b\xi^c\right\}\ea
This is the approximation underlying our discussion
in Section 2. In fact,
if we insert this truncation into the exact evolution equation
(\ref{AA.13}), we obtain precisely eq. (\ref{3}) whose structure
we explained already.

Moreover, a generic functional
$\Gamma_k[A,\bar A]$ can be decomposed according to
\be\label{A.25}
\Gamma_k[A,\bar A]=\bar\Gamma_k[A]+\frac{1}{2\alpha}\int d^dx
[D_\mu(\bar A)(A_\mu-\bar A_\mu)]^2+\hat\Gamma_k^{\rm gauge}[A,\bar A]
\ee
where $\bar\Gamma_k$ is defined by equating the two-gauge fields:
$\bar\Gamma_k[A]\equiv \Gamma_k[A,A]$. The remainder $\Gamma_k[A,\bar
A]-\bar\Gamma_k[A]$ is further decomposed in the classical
gauge-fixing term plus a correction to it, $\hat\Gamma_k^{\rm gauge}$.
Note that $\hat\Gamma_k^{\rm gauge}[A,A]=0$ for equal gauge fields.
 $\bar\Gamma_k
[A]$ is a gauge-invariant functional of $A_\mu$ and $\Gamma_k[A,\bar A]$
is invariant under a simultaneous gauge transformation of $A$
and $\bar A$.
In the examples of the following sections
we make the further approximation of neglecting
quantum corrections to the gauge fixing term by setting
$\hat\Gamma_k^{\rm gauge}=0$.
The important question is whether this truncation is consistent
with the Ward-Takahashi identities (\ref{AA.16}) and the
$\bar A$-derivative (\ref{AA.20}), respectively. If we insert
(\ref{A.23})-(\ref{A.25}) into (\ref{AA.16}) ,
we find that $\bar\Gamma_k$ drops out from the LHS
of this equation. We are left with a condition for $\hat\Gamma_k
^{\rm gauge}$:
\be\label{A.27}
-\frac{1}{g}\int d^dx\frac{\delta\hat\Gamma^{\rm gauge}}{\delta\bar
A_\mu^a(x)}(D_\mu(A)\xi)^a(x)=\Delta_k^{(\rm BRS)}\ee
The anomaly $\Delta^{(\rm BRS)}_k$ (\ref{AA.16a})
vanishes for $k\to0$ but is non-zero for $k>0$. Our
approximation $\hat\Gamma_k^{\rm gauge}\equiv0$ is consistent
provided these terms can be neglected. We note that the traces
implicit in (\ref{A.27}) are related to higher loop effects.
Beyond a loop approximation our neglection of $\hat\Gamma^{\rm gauge}$
is a non-trivial assumption. We emphasize that because of its
gauge invariance the functional $\bar\Gamma_k[A]$ does not appear
on the LHS of the Ward identities. Therefore the Ward identities
do not imply any further condition for $\bar\Gamma_k$. This means
that, within the approximations made, we may write down any
ansatz for $\bar\Gamma_k$ as long as it is gauge-invariant.
Similar remarks apply to the identity
(\ref{AA.20})
for the $\bar A$-dependence.

\renewcommand{\theequation}{4.\arabic{equation}}
\setcounter{equation}{0}
\section*{4 Evolution of the Nonabelian Gauge Coupling}
The exact evolution equation
is a nonlinear differential equation for a function of infinitely
many
variables.
There seems to be little hope for finding closed-form solutions.
The successful
use of this equation therefore depends crucially on the existence of an
appropriate
approximation scheme.
This will consist in a truncation of the infinitely
many invariants characterizing $\Gamma_k$ to a finite number.
If one makes an ansatz for $\Gamma_k$ which
contains only finitely many parameters (depending on $k$) and
inserts it into (\ref{3}), the functional differential equation
reduces to a set of coupled ordinary differential equations for
the parameter functions.
The truncation
should be chosen in such a way that it encapsulates the essential
physics in an ansatz as simple as possible. In a second step one
has to verify that upon including more terms in the truncation
the results do not change significantly any more.

In this section we demonstrate the
practical use of our equation by computing
approximately
the running of the nonabelian gauge coupling
of pure Yang-Mills theory with gauge group SU(N)
in arbitrary dimension $d$ \cite{ex}.
In order to approximate the
solution $\Gamma_k[A,\bar A]$ of (\ref{3}) by a
functional with
at most second derivatives we make the ansatz
\be\label{7.2}
\Gamma_k[A,\bar A]=\int d^dx\left\lbrace \frac{1}{4}Z_{Fk}
\ F^a_{\mu\nu}(A)F_{\mu\nu}^a(A)+
          \frac{Z_{Fk}}{2\alpha_k}[D_\mu[\bar A](A_\mu-\bar
A
_\mu)]^2\right\rbrace\ee
We want to determine the running of $Z_{Fk}$ from the flow equation.
The truncation (\ref{7.2}) leads to the Hessian
\be\label{7.3}
\frac{\delta^2\Gamma_k[A,\bar A]}{\delta A_\mu^a(x)\delta A_\nu^b(x')}=
Z_{Fk}\left\lbrace {\cal D}_{\rm T}[A]_{\mu\nu}+D_\mu
[A]D_\nu[A]-
\frac{1}{\alpha_k}D_\mu[\bar A]D_\nu[\bar A]\right\rbrace^{ab}
\delta(x-x')\ee
where
$({\cal D}_T)_{\mu \nu}\equiv
 - D^2\delta_{\mu\nu}+2i\bar g F_{\mu%
\nu}$~with the color matrix $F$~in the adjoint representation.
($\bar{g}$~denotes the bare gauge coupling.)
In the following we neglect the running of $\alpha_k$ and
restrict our discussion to $\alpha_k=1$. Thus
\be\label{7.4}
\frac{\delta^2}{\delta A^2}\Gamma_k[A,\bar A]\Bigr
\vert_{\bar A=A}=Z_{Fk}{\cal D}_{\rm T}
(A)\ee
and the evolution equation reads for $\bar A=A$:
\begin{eqnarray}\label{7.5}
\frac{\partial}{\partial t}\Gamma_k[A,A]&=&\frac{\partial Z_{Fk}}{\partial t}
\int d^dx\frac{1}{4}F_{\mu\nu}^{a}F_{\mu\nu}^{a}\nonumber\\
&=&\frac{1}{2}{\rm Tr}\left[\left(\frac{\partial}{\partial t}R_k({      \cal
D}_{\rm T}
)\right)\left(Z_{Fk}{\cal D}_{\rm T}+R_k({\cal D}_{\rm
T})\right)^{-1}\right]
\nonumber\\
&&-{\rm Tr}\left[\left(\frac{\partial}{\partial t}R_k({\cal D}_{\rm
S})\right)
\left({\cal D}_{\rm S}+R_k({\cal D}_{\rm
S})\right)^{-1}\right]\end{eqnarray}
Here ${\cal D}_{\rm T}$ and ${\cal D}_{\rm S}\equiv -D^2$
     depend on $A_\mu$ now. For mathematical convenience we chose
$\Delta={\cal D}_T$ for the cutoff operator.
The function
$R_k$ is defined
with $Z_k=Z_{Fk}$ in the first trace on the
RHS of (\ref{7.5}) (gluons) and
with $Z_k=1$ in the second trace (ghosts).
In order
to
determine $\partial Z_{Fk}/\partial t$ it is sufficient to extract
the term proportional
to the invariant $F_{\mu\nu}^{a}F_{\mu\nu}^{a}$~from the traces
on the RHS of (\ref{7.5}).
This can be done by using standard heat-kernel
techniques or by inserting a simple field configuration
on both sides of the equation for which the traces can be calculated
easily. In either case one finds
for $d>2$ \cite{ex}
\begin{eqnarray}\label{7.12}
\frac{\partial}{\partial t} Z_{Fk}=&-&2N\left(1-\frac{d}{24}\right)
\frac{v_{d-2}}{\pi}\bar g^2
\int^\infty_0 dx\ x^{\frac{d}{2}-2}\frac{d}{dx}\frac{\partial_t R_k(x)}
{Z_{Fk} x+R_k(x)}\nonumber\\
&-&\frac{1}{6}N\frac{v_{d-2}}{\pi}\bar g^2\int^\infty_0 dx
\  x^{\frac{d}{2}-2}\frac{d}{dx}\frac{\partial_tR_k(x)}{x+R_k(x)}
\ \equiv\ \bar g^2\  b_d\  k^{d-4}                       \end{eqnarray}
with $v_d\equiv\left[ 2^{d+1}\pi^{d/2}\Gamma(d/2)\right]^{-1}$.
The second integral is due to the trace containing ${\cal D}_{\rm  S}$
with $Z_k=1$ in $R_k(x)$.
Introducing the dimensionless, renormalized gauge coupling
\be\label{7.13}
g^2(k)=k^{d-4}\ Z^{-1}_{Fk}\ \bar g^2
\ee
the associated beta function reads
\be\label{7.14}
\beta_{g^2}\equiv\frac{\partial}{\partial t} g^2(k)
=(d-4)g^2+\eta_F\ g^2=(d-4) g^2-b_d\ g^4.
\ee
where $\eta_F\equiv -\partial_t \ln Z_{Fk}$~denotes
the anomalous dimension.
For $d=4$ the result for the
running of $g^2(k)$ becomes universal, i.e., $b_4$
is
independent of the precise form of
the cutoff function $R_k(x)$, only its
behavior
for $x\to0$ enters in (\ref{7.12}).
One obtains, with $\lim_{x\to 0} R_k=Z_{Fk}
k^2$
for the first term
in (\ref{7.12}) and lim$_{x\to0}R_k=k^2$ for the second term,
\be\label{7.15}
b_4=\frac{N}{24\pi^2} \Big[11-5\eta_F\Big]
\ee
In lowest order in $g^2$ we can
neglect $\eta_F$ on the RHS of (\ref{7.15})
and
obtain the standard
perturbative one-loop $\beta$-function. More generally,
one finds for $\eta_F$ the equation
$
\eta_F=-g^2b_d(\eta_F)$,
which has, for $d=4$, the nonperturbative solution
\be\label{7.17}
\eta_F=-\frac{11N}{24\pi^2}g^2\left[1-\frac{5N}{24\pi^2}g^2\right]%
                                                            ^{-1}\ee
The resulting $\beta$-function can be expanded for small $g^2$
\begin{eqnarray}\label{7.18}
\beta_{g^2}&=&-\frac{11N}{24\pi^2}g^4\left[1-\frac{5N}{24\pi^2}g^2\right]^{-1}
\nonumber\\
&=&-\frac{22N}{3}\frac{g^4}{16\pi^2}-\frac{220}{9}N^2\frac{g^6}{(16\pi^2)^2}
-...
\end{eqnarray}
Comparing with the standard perturbative two-loop expression
\be\label{7.19}
\beta^{(2)}_{g^2}=-\frac{22N}{3}\frac{g^4}{16\pi^2}-\frac{204}{9}
N^2\frac{g^6}{(16\pi^2)^2}\ee
we find a surprisingly good agreement even for the two-loop coefficient. The
missing 7 \% in the coefficient of the $g^6$-term in $\beta_{g^2}$ should be
due to our truncations.

For arbitrary $d$ we introduce
the constants $l^d_{\NA}$ and
$l^d_{\NA\eta}$
by
\be\label{7.20}
b_d=\frac{44}{3}N\ v_d\ l^d_{\NA}-
     \frac{20}{3}N\ v_d\ l^d_{\NA\eta}\ \eta_F         \ee
They are normalized such that in 4 dimension
$l^4_{\NA}=1, l^4_{\NA\eta}=1$ for any choice of the cutoff function.
For $d$~different from 4 they are not universal. If we use the
exponential cutoff function (\ref{5}) they read for $d>2$:
\begin{eqnarray}\label{7.22}
l^d_{\NA}&=&-\frac{1}{88}(26-d)%
(d-2)k^{4-d}\int^\infty_0dx\ x^{\frac{d}{2}-2}
\frac{d}{dx}\frac{d}{dt}\ln P\nonumber\\
&=&\frac{(26-d)(d-2)}{44}\ n^{d-4}_1\end{eqnarray}
\begin{eqnarray}\label{7.23}
l^d_{NA\eta}&=&-\frac{1}{40}(24-d)(d-2)k^{4-d}\int^\infty_0dx\ x^{\frac{d}{2}-2}
\frac{d}{dx} \frac{P-x}{P}\nonumber\\
&=&\frac{(24-d)(d-2)}{40}\ l^{d-2}_1\end{eqnarray}
It is remarkable that the $\beta$-function
for the dimensionful, renormalized
coupling $g^2_R=g^2k^{4-d}$ vanishes precisely in the critical
string dimension $d=26$.
More explicitly, one has
\begin{eqnarray}\label{7.25}
n^{d-4}_1&=&
-\frac{1}{2}k^{4-d}\int^\infty_0dx\ x^{\frac{d}{2}-2}
\frac{\partial}{\partial t}\frac{dP/dx}{P}\nonumber\\
&=&-\int^\infty_0dy\ y^{\frac{d}{2}-2}e^{-y}(1-y-e^{-y})
(1-e^{-y})^{-2}>0\end{eqnarray}
\be\label{7.24}
l^{d-2}_1=\Gamma\left(\frac{d-2}{2}\right)\ee
The evolution equation for the running dimensionless renormalized gauge coupling
$g$
in arbitrary dimension
\be\label{7.26}
\frac{\partial g^2}{\partial t}=\beta_{g^2}=(d-4)g^2-\frac{44N}{3}
v_d\ l^d_{\NA}\ g^4\left[1-\frac
{20N}{3} v_d\ l^d_{\NA\eta}\ g^2\right]^{-1}\ee
has the general solution (for $d\not=4$)
\be\label{7.27}
\frac{g^2(k)}{[1+a_2g^2(k)]^\gamma}=C\left[\frac{k}{k_0}\right]^{d-4}\ee
with
\begin{eqnarray}\label{7.28}
a_1&=&\frac{44N\ v_d\ l^d_{\NA}}{3(4-d)}\nonumber\\
a_2&=&a_1-\frac{20N}{3}{v_d}\ l^d_{\NA\eta}\nonumber\\
\gamma&=&a_1/a_2\end{eqnarray}
and
\be\label{7.29}
C=\frac{g^2(k_0)}{[1+a_2g^2(k_0)]^\gamma}\ee
The nonabelian Yang-Mills theory is asymptotically free for $d\leq 4$ with a
``confinement scale''
$\Lambda^{(d)}_{\conf}$, where $\beta_{g^2}$ diverges
\be\label{7.30}
\Lambda^{(d)}_{\conf}=%
                     \left[\frac{Ca_1^\gamma}{(a_1-a_2)^{\gamma-1}}\right]^{%
\frac{1}{4-d}}
k_0\ee
At this scale our truncation gives  no quantitatively reliable
results any more since $\eta_F$ diverges and the choice $Z_k=Z_{Fk}$ in $R_k$
becomes
inconvenient. Indeed, $Z_{Fk}$ may vanish for some scale $k_{cf}>0$, whereas
$Z_k$ should always remain strictly positive. A possible smoother definition
in the region of rapidly varying $Z_{Fk}$ could be $Z_\Lambda=Z_{F\Lambda}$ for
$k=\Lambda$, and
$\partial_tZ_k=-\eta_F(1+\eta_F^2)^{-1}Z_k$ for $k<\Lambda$.
This modification does not influence
the one and two loop $\beta$-function. It guarantees, however, that $Z_k$
remains always
strictly positive. Now the $\beta$ function does not diverge for any finite
value
of $g^2$ and the confinement scale can always be associated with the scale where
$g^2$
diverges or $Z_{Fk}$ vanishes. This scale is slightly lower than (\ref{7.30}).
The ``one-loop'' confinement
scale obtains from (\ref{7.30})
for $l^d_{\NA\eta}\to0,\ a_2\to a_1,\ \gamma\to1$
\be\label{7.31}
\Lambda_{\conf}^{(d)}
=\left[\frac{44Nv_dl^d_{\NA}}{3(4-d)}g^2(k_0)\left(1+
\frac{44Nv_dl^d_{\NA}}%
{3(4-d)}g^2(k_0)\right)^{-1}\right]^{\frac{1}{4-d}}k_0\ee
and corresponds as usual to a diverging gauge coupling. We observe that $\Lambda
^{(d)}_{\conf}$ (\ref{7.30}) %
                            is always higher than the ``one-loop'' result
(\ref{7.31}) (for given $k_0$ and $g^2(k_0))$. We therefore consider the scale
(\ref{7.31}) as a lower bound for the confinement scale.

For $4<d<24$ the
$\beta$
function (\ref{7.26}) has an ultraviolet stable fixpoint separating the
confinement
phase for strong coupling (with a confinement scale given by the analog of
(\ref{7.30}) for negative $a_1$ and $a_2$) from the infrared free weak coupling
phase. We note that there is no confinement phase for $d>26$.

\renewcommand{\theequation}{5.\arabic{equation}}
\setcounter{equation}{0}
\section*{5 Chern-Simons Theory}

As a second example we now turn to pure Chern-Simons theory
in 3 dimensions. This is an interesting
theory from many points of view. It can be used to give a
path-integral representation of knot and link invariants
\cite{wi} and to understand many properties of
2-dimensional conformal field theories \cite{wi,conf}. Being a
topological field theory the model has no propagating degrees of
freedom. Canonical quantization yields a
Hilbert space with only finitely many physical states which can be
related to the conformal blocks of (rational) conformal field
theories. Perturbative covariant quantization
\cite{alv,gia,fal,mar,shif} shows that the theory is not only
renormalizable but even ultraviolet finite. It is
remarkable that despite this high degree of "triviality" the
theory produces nontrivial radiative corrections.
One-loop effects were found \cite{pi,wi} to lead to a
renormalization of the parameter $\kappa$ which multiplies the
Chern-Simons 3-form in the action,
\be   \label{1}
     S_{\CS}[A]=i\kappa \ \frac{g^2}{8\pi} \int d^3\!x
     ~\varepsilon_{\alpha\beta\gamma} \: [A^a_\alpha\,
     \partial_{\beta}\! A^a_\gamma + \frac{1}{3}g
     f^{abc} A^a_{\alpha} A^b_{\beta} A^c_{\gamma}]
\ee
A variety of gauge invariant regularization methods, including
spectral flow arguments based upon the $\eta$-invariant, predict
a finite difference between the bare and the renormalized value
of
$\kappa$:
\be          \label{2}
     \kappa_{\ren}=\kappa_{\bare}+\sign(\kappa)~T(G)
\ee
Here $T(G)$ denotes the value of the quadratic Casimir operator
of the gauge group $G$ in the adjoint representation. It is
normalized such that $T(SU\!(N))=N$. The shift of $\kappa$ has
a natural relation to similar shifts in the Sugawara construction
of 2-dimensional conformal field theories. On the other hand, in
standard renormalization theory a relation of the type (\ref{2}) is
rather unusual, and there has been some controversy in the literature
about the correct interpretation of
eq. (\ref{2}). Following ref. \cite{cspp}
we shall investigate this problem in the context of the effective
average action now.

Let us try to find an approximate solution
of the initial value problem (\ref{3}) with (\ref{4})
for the classical Chern-Simons action (\ref{1}).
We work on flat euclidean
space and allow for an arbitrary semi-simple, compact gauge group
$G$.
We use a truncation of the
form \cite{cspp}
\ba   \label{6}
     \Gamma_k[A,N,\bar{A}] &=&
     i\kappa(k)~\frac{g^2}{4\pi}~I[A]+
     \kappa(k)~\frac{g^2}{8\pi}    \int d^3\!x\,\Bigl\{iN^a
     D_{\mu}^{ab}[\bar{A}]\, (A^b_{\mu}-\bar{A}^b_{\mu})
              \nonumber             \\
     &&-i(A^a_{\mu}-\bar{A}^a_{\mu})D^{ab}_{\mu}[\bar{A}]~N^b
      +\alpha\, \kappa(k)
     \frac{g^2}{4\pi}N^aN^a\Bigr\}
\ea
with
\be
     I[A] \equiv \frac{1}{2} \int
     d^3\!x~\varepsilon_{\alpha\beta\gamma}
     ~[A^a_{\alpha}\,\partial_\beta\! A^a_\gamma +
     \frac{1}{3}g f^{abc}
     A^a_{\alpha}A^b_{\beta}A^c_{\gamma}]
\ee
The first term on the RHS of (\ref{6}) is the Chern-Simons action, but
with a scale-dependent prefactor. In the second term we
introduced an auxiliary field $N^a(x)$ in order to linearize the
gauge fixing term. By eliminating $N^a$ one recovers the
classical, $k$-independent background gauge fixing term
     $\frac{1}{2\alpha}(D_{\mu}[\bar{A}](A_{\mu}-\bar{A}_{\mu}))^2$.
As we discussed in Section 3,
also the gauge fixing term could in principle change its form
during the evolution, but this effect is neglected here.

For $k\rightarrow \infty$, and upon eliminating $N^a$, the ansatz
(\ref{6}) reduces to (\ref{4}) with the identification
     $\kappa(\infty) \equiv \kappa_{\bare}$.
We shall insert (\ref{6}) into the evolution equation and from the
solution for the function $\kappa(k)$ we shall be able to
determine the renormalized parameter
     $\kappa(0) \equiv \kappa_{\ren}$.
We have to project the traces on the RHS of (\ref{3}) on the subspace
spanned by the truncation (\ref{6}). This means that we have
to extract only the term proportional to $I[A]$ and to compare
the coefficients of $I[A]$ on both sides of the equation. In the
formalism with the auxiliary field, $\Gamma^{(2)}_k$ in (\ref{3})
denotes the matrix of second functional derivatives with respect
to both $A^a_\mu$ and $N^a$, but with $\bar{A}^a_\mu$ fixed.
Setting
$\bar{A}=A$
after the variation, one obtains
 \ba     \label{8}
     \delta^2\Gamma_k[A,N,A]&= & i\kappa(k) \frac{g^2}{4\pi}
     \int d^3\!x~\Big\{\delta\! A^a_{\mu}\, \varepsilon_{\mu \nu
     \alpha} D^{ab}_{\alpha}\, \delta\! A^b_{\nu}+\delta\!
     N^aD^{ab}_{\mu}\,\delta\! A^b_{\mu}   \nonumber \\
                                      && -\delta\!
     A^a_{\mu}D^{ab}_{\mu}N^b\Big\}
     +\alpha \ (\kappa(k) \frac{g^2}{4\pi})^2 \int
     d^3x~\delta\! N^a~\delta\! N^a
 \ea
In order to facilitate the calculations we introduce three
4$\times$4 matrices $\gamma_\mu$ with matrix elements
$(\gamma_\mu)_{mn}$, $m$=($\mu$,4)=1,...,4, etc., in the
following way \cite{shif}:
 \be  \label{9}
 (\gamma_\mu)_{\alpha\beta}=\varepsilon_{\alpha\mu\beta},\ %
     (\gamma_\mu)_{4\alpha}=-(\gamma_\mu)_{\alpha 4}
     =\delta_{\mu\alpha},\ %
     (\gamma_\mu)_{44}=0
 \ee
If we combine the gauge field fluctuation and the auxiliary field
into a 4-component object $\Psi^a_m \equiv (\delta
A^a_{\mu},\delta N^a)$ and choose the gauge $\alpha=0$, we find
\be
     \delta^2\Gamma_k[A,N,A] = i\kappa(k) \frac{g^2}{4\pi}
     \int d^3\!x ~\Psi^a_m(\gamma_\mu)_{mn} D^{ab}_{\mu}
     \Psi^b_n
\ee
so that in matrix notation
\be  \label{11}
     \Gamma^{(2)}_k=i\kappa(k) \  \frac{g^2}{4\pi} \D
\ee
Clearly $\D \equiv \gamma_{\mu}D_{\mu}$ is reminiscent of a Dirac
operator. In fact, the algebra of the $\gamma$-matrices is similar
to the one of the Pauli matrices:  \be
\gamma_{\mu}\gamma_{\nu}=-\delta_{\mu\nu}
+\varepsilon_{\mu\nu\alpha}\gamma_{\alpha}  \ee
Because $\gamma^+_{\mu}=-\gamma_{\mu}, \ \ \D$ is hermitian. Its
square reads
\be
     \D^2=-D^2-ig\ ^*\!F_\mu\gamma_{\mu}
\ee
where  \be
      ^*\!F_{\mu} \equiv \frac{1}{2}
     \varepsilon_{\mu\alpha\beta}F_{\alpha\beta}     \ee
is the dual of the field strength tensor.
Because $\D^2$ is essentially the
covariant laplacian, it is the natural candidate for the cutoff
operator $\Delta$. With this choice, and\be
c \equiv \frac{g^2}{4\pi}                  \ee
the evolution equation (\ref{3})
reads at $\bar{A}=A$:
\ba         \label{13}
     ic ~k \frac{d}{dk}\kappa(k)~I[A] &=& \frac{1}{2} \Tr
 \left[\left(ic\kappa \D+R_k(\D^2)\right)^{-1} k \frac{d}{dk}R_k(\D^2)%
\right] \nonumber   \\  &&-  \Tr
\left[\left(-D^2+R_k(-D^2)\right)^{-1} k \frac{d}{dk}R_k(-D^2)%
 \right]
\ea
The second trace on the RHS of (\ref{13}) is due to the ghosts.
It is manifestly
real, so it cannot match the purely imaginary $i I[A]$ on the LHS
and can be
omitted therefore. For the same reason we may replace the first
trace by $i$ times its imaginary part:
\be \label{14}
     k \frac{d}{dk} \kappa(k)\, I[A] = - \frac{1}{2}
     \kappa(k)\,\Tr\left[\, \D\ \left(c^2 \kappa^2%
     \D^2 +R^2_k(\D^2)\right)^{-1}
     k \frac{d}{dk}R_k(\D^2)\right]
     +\cdots
\ee
The trace in (\ref{14}) involves an integration over spacetime, a
summation over adjoint group indices, and a ``Dirac trace". We
shall evaluate it explicitly in the next section. Before turning
to that let us first look at the general structure of eq. (\ref{14}). In
terms of the (real) eigenvalues $\lambda$ of $\D$ eq. (\ref{14}) reads
\be                           \label{15}
     \frac{d\kappa(k)}{dk^2}~I[A]=-\frac{1}{2}\kappa(k)
     \sum_{\lambda}\frac{\lambda}{c^2\kappa^2(k)
     \lambda^2+R^2_k(\lambda^2)} \cdot
     \frac{dR_k(\lambda^2)}{dk^2}
\ee
where we switched from $k$ to $k^2$ as the independent variable.
We observe that the sum in (\ref{15}) is related to a regularized form
of the spectral asymmetry of $\DD$.

An approximate solution for $\kappa(k)$ can be obtained by
integrating both sides of eq. (\ref{15}) from a low scale $k^2_0$ to a
higher scale $\Lambda^2$ and approximating $\kappa(k) \simeq
\kappa(k_0)$ on the RHS. This amounts to ``switching off''
the renormalization group improvement.
The result is
\be                       \label{16}
     [\kappa(k_0)-\kappa(\Lambda)]~I[A] = \frac{1}{2}
     \kappa(k_0)
     \sum_{\lambda}
                 \int^{\Lambda^2}_{k_0^2} dk^2
  \ \frac{dR_k(\lambda^2)}{dk^2} \cdot
     \frac{\lambda}{c^2\kappa^2(k_0)
     \lambda^2+R^2_k(\lambda^2)}
\ee
Upon using $R_k$ as the variable of integration one arrives at
\be   \label{17}
     [\kappa(k_0)-\kappa(\Lambda)]~I[A] =
     \frac{1}{2c}\ \sign(\kappa(k_0)) \sum_{\lambda}~
     \sign(\lambda)\, G(\lambda;k_0,\Lambda)
\ee
with
\be             \label{18}
     G(\lambda;k_0,\Lambda) \equiv \arctan\left[
     c\,|\kappa(k_0)\lambda|\,
     \frac{R_\Lambda(\lambda^2)-R_{k_0}(\lambda^2)}
     {c^2\kappa(k_0)^2\lambda^2 +
     R_{\Lambda}(\lambda^2)~R_{k_0}(\lambda^2)} \right]
\ee
Recalling the properties of $R_k$ we see that in the spectral
sum (\ref{17}) the contributions of eigenvalues $|\lambda| \ll k_0$
and $|\lambda|\gg\Lambda$ are strongly suppressed, and only the
eigenvalues with $k_0 < |\lambda| < \Lambda$ contribute
effectively. Ultimately we would like to perform the limits $k_0
\rightarrow 0$ and $\Lambda \rightarrow \infty$. In this case the
sum over $\lambda$ remains without IR and UV regularization. This
means that if we want to formally perform the limits $k_0
\rightarrow 0$ and $\Lambda \rightarrow \infty$ in eq. (\ref{17}), we
have to introduce an alternative regulator. In order to make
contact with the standard spectral flow argument \cite{wi} let us
briefly describe this procedure. We avoid IR divergences by
putting the system in a finite volume and imposing boundary
conditions such that there are no zero modes. In the UV we
regularize with a zeta-function-type convergence factor
$|\lambda/\mu|^{-s}$ where $\mu$ is an arbitrary mass
parameter. Thus the spectral sum becomes
\be
     \lim_{s \rightarrow 0}\  \sum_{\lambda} \sign(\lambda)\,
     \left|\lambda / \mu\right|^{-s} G(\lambda;
     k_0,\Lambda)
\ee
Now we interchange the limits $k_0 \rightarrow 0$, $\Lambda
\rightarrow \infty$ and $s \rightarrow 0$. By construction, only
finite $(|\lambda| \leq \mu)$ and nonzero eigenvalues
contribute. For such $\lambda$'s we have
$G(\lambda; 0,\infty)=\pi/2$ irrespective of the precise form of
$R_k$. Therefore (\ref{17}) becomes
\be
     [\kappa(0)-\kappa(\infty)]~I[A]=
     \frac{2\pi^2}{g^2}~\sign(\kappa(0)) ~\eta[A]
\ee
where  \be  \eta[A] \equiv \lim_{s \rightarrow 0} \frac{1}{2}
\sum_{\lambda} \sign(\lambda)~|\lambda/\mu|^{-s} \ee    is the
eta-invariant. If we insert the known result \cite{wi}
\be       \eta[A]=(g^2/2\pi^2)~T(G)~I[A]   \ee          we recover
eq.(\ref{2}):
$
     \kappa(0)=\kappa(\infty)+\sign(\kappa(0))~T(G)
$.
Obviously $R_k$ has
dropped out of the calculation. The
parameter $\kappa$ is universal: it does not depend on the form
of the IR cutoff.

\renewcommand{\theequation}{6.\arabic{equation}}
\setcounter{equation}{0}
\section*{6 Evolution of the Chern-Simons Parameter}

Next we turn to an explicit evaluation of the trace in eq. (\ref{14})
which keeps the full $k$-dependence of
$\kappa$ on the RHS, i.e., the
renormalization group improvement. To start with
we use the constant cutoff
\footnote{As the Faddeev-Popov
ghosts do not contribute to the effect under
consideration we may set $Z_k=1$~also in the cutoff
for the gauge field.}
                           $R_k=k^2$ for which eq. (\ref{14}) assumes
the form
\be      \label{22}
     \frac{d}{dk^2}\kappa(k)~I[A]
=-\frac{1}{2c^2\kappa(k)}%
 \,\Tr\left[\, \D\left(\D^2+l(k)^2\right)^{-1}\right]
\ee
where
\be           \label{23}
     l(k) \equiv \frac{k^2}{c~|\kappa(k)|}
\ee
If we extract from the
trace the term quadratic in $A$ and linear in the external
momentum and equate the coefficients of the $A\,\partial \!
A$-terms on both sides of (\ref{22}) we obtain
\be \label{24}
     \frac{d\kappa(k)}{dk^2} \int d^3\!x
     ~\varepsilon_{\alpha\beta\gamma}
     ~A^a_{\alpha}\,\partial_{\beta}\!A^a_{\gamma}=
     -\frac{g^2T(G)}{c^2\kappa(k)} \int d^3\!x\,
     \varepsilon_{\alpha\beta\gamma}\,
     A^a_{\alpha}\Pi_k(-\partial^2)
     \partial_{\beta}\!A^a_{\gamma}+O(A^3)
\ee
The function $\Pi_k$ is given by the Feynman parameter integral
\be  \label{25}
     \Pi_k(q^2)=8 \int_0^1 dx~x(1-x) \int
     \frac{d^3p}{(2\pi)^3} \,
     \frac{q^2}{[p^2+l^2+x(1-x)q^2]^3}
\ee
Expanding $\Pi_k(-\partial^2)= \Pi_k(0)-\Pi'_k(0)\partial^2+
...$, we see that only for the term with $\Pi_k(0)$ the number of
derivatives on both sides of eq.(24) coincides. Therefore one
concludes that
\be   \label{26}
     \frac{d\kappa(k)}{dk^2}= - \frac{g^2
     T(G)}{c^2\kappa(k)} \  \Pi_k(0)
\ee
where $\Pi_k(0)$ depends on $\kappa(k)$ via
(\ref{23}). Equation (\ref{26}) is
the renormalization group equation for $\kappa(k)$ which we
wanted to derive. Formally it is similar to the evolution
equation in Section 4 or the ones of the abelian Higgs
model \cite{ahm}. The special features of Chern-Simons
theory, reflecting its topological character, become obvious when
we give a closer look to the function $\Pi_k(q^2)$. Assume we fix
a non-zero value of $k$ $(l\neq 0)$ and let $q^2 \rightarrow 0$
in (\ref{25}). Because the $l^2$-term prevents the $p$-integral from
becoming IR divergent, we may set $q^2=0$ in the denominator, and
we conclude that the integral vanishes $\sim q^2$. This means
that the RHS of (\ref{26}) is zero and
that $\kappa(k)$ keeps the same value for all strictly positive
values of $k$.
However, $\Pi_k(0)$ really vanishes
only for $k>0$. If we set $l=0$ in (\ref{25}) we cannot conclude
anymore that $\Pi_k \sim q^2$, because in the region $p^2
\rightarrow 0$ the term $x(1-x)q^2$ provides the only IR cutoff
and may not be set to zero in a naive way. In fact, $\Pi_k(0)$
has a $\delta$-function-like peak at $k=0$. To see this, we first
perform the integrals in (\ref{25}):
\be             \label{27}
     \Pi_k(q^2)=\frac{1}{\pi}\left[ \frac{1}{2|q|} \arctan
     \left(\frac{|q|}{2|l|}\right)-\frac{|l|}{q^2+4l^2}
     \right]
\ee
As $q^2$ approaches zero, this function develops an increasingly
sharp maximum at $l=0$. Integrating (\ref{27}) against a smooth test
function $\Phi(l)$ it is easy to verify that
\be                                 \label{28}
     \lim_{q^2 \rightarrow 0} \int_0^{\infty} dl~
     \Phi(l)~\Pi_k(q^2) = \frac{1}{4\pi} \Phi(0)
\ee
This means that on the space of even test functions \be   \lim_{q^2
\rightarrow 0}\Pi_k(q^2)=\frac{1}{2\pi} \delta(l)    \ee
Even though the value
of $\kappa(k)$ does not change during almost the whole evolution
from $k=\infty$ down to very small scales, it performs a finite
jump in the very last moment of the evolution, just before
reaching $k=0$. This jump can be calculated in a well-defined
manner by integrating (\ref{26}) from $k^2=0$ to $k^2=\infty$:
\be  \label{29}
     \kappa(0)-\kappa(\infty)= 4\pi~T(G)~\lim_{q^2
     \rightarrow 0} \int_0^{\infty} dl~\sign(\kappa(l))
     \cdot \left[1-c \, l \frac{d}{dk^2} |\kappa(k)|\right]^{-1}
     \Pi_k(q^2)
\ee
The term $\sim d|\kappa|/dk^2$ is a Jacobian factor which is due
to the fact that $l$ depends on $\kappa(k)$. This factor is the only
remnant of the $\kappa(k)$-dependence of the RHS of the evolution
equation. As we saw in Section 4,
this dependence of the RHS on the running couplings is
the origin of the renormalization group improvement.
If we use (\ref{28}) in (\ref{29}),
$l~d|\kappa|/dk^2$ is set to zero and we find
\be   \label{30}
     \kappa(0)=\kappa(\infty)+\sign(\kappa(0))~T(G),
\ee
which is precisely the 1-loop result. It is straightforward to
check that the shift (\ref{30}) is independent of the choice for $R_k$.

It is quite instructive to compare
the situation in Chern-Simons theory with what we found for
ordinary Yang-Mills theory in Section 4.
Like $\kappa$, also the gauge coupling in QCD$_4$~%
is a universal quantity. Its running is governed by a
$R_k$-independent $\beta$-function which leads to a logarithmic
dependence on the scale $k$. The Chern-Simons parameter $\kappa$,
on the other hand, does not run at all between $k=\infty$ and any
infinitesimally small value of $k$. Only at the very end of the
evolution, when $k$ is very close to zero, $\kappa$ jumps by a
universal, unambiguously calculable amount $\pm T(G)$. Though
surprising in comparison with non-topological theories, this
feature is precisely what one would expect if one recalls the
topological origin of a non-vanishing $\eta$-invariant \cite{wi}.
If $\eta[A]\neq 0$ for a fixed gauge field $A$,
some of the low lying eigenvalues of $\D[A]$ must have
crossed zero during the interpolation from $A=0$ to $A$. However,
this spectral flow involves only that part of the spectrum which,
in the infinite volume limit, is infinitesimally close to zero.

The jump of $\kappa$~is also the resolution to the following apparent
paradox.
The effective average
action $\Gamma_k$ is closely related to a continuum version of
the block-spin action of lattice systems.
Block-spin transformations can be iterated, and when we
have already constructed $\Gamma_{k_1}$ at a certain scale $k_1$
we may view $\Gamma_{k_1}$ as the ``classical" action for the next
step of the iteration, in which an integral over
$\exp{(-\Gamma_{k_1})}$ has to be performed. Trying
to understand the
shift (\ref{2}) from a renormalization group point of view, we are
confronted with the following puzzle. Because
$S_{\CS}$ is not invariant under large gauge transformations,
$\exp{(-S_{\CS})}$ is single valued only if $\kappa \in {\bf Z}$.
If there is a {\it continuous}
interpolation between $\kappa(\infty)$ and $\kappa(0)$ a
nontrivial shift means that there are intermediate scales at
which $\kappa$ cannot be integer.
This suggests that
$\kappa_{\ren}=\kappa_{\bare}$,
because there should be an
inconsistency if we try to do the next blockspin
transformation starting from a
multivalued Boltzmann factor $\exp{(-\Gamma_{k_1})}$.
It is clear now that this argument does not apply precisely
because the trajectory from
$\kappa(\infty)$~to $\kappa(0)$
is not continuous.

Another unusual feature of Chern-Simons theory is the absence
of any renormalization group improvement beyond the 1-loop
result. This should be contrasted with the running of
$g$ in QCD$_4$ where the truncation of Section 4
leads to a nonperturbative $\beta$-function
involving arbitrarily high powers of $g$. We emphasize that our
evolution equation with the truncation (\ref{6}) potentially goes
far beyond a 1-loop calculation. It is quite remarkable therefore
that in Chern-Simons theory all higher contributions vanish.
From the discussion following eq. (\ref{29}) it is clear that
this is again due to the unusual discontinuous behavior
of $\kappa$ which reflects the topological field theory
nature of the model. While it
is not possible to translate a ``nonrenormalization theorem"
for a given truncation into a statement about the
nonrenormalization at a given number of loops, our results
point in the same direction as ref.~\cite{gia} where the
absence of 2-loop corrections was proven.

\renewcommand{\theequation}{7.\arabic{equation}}
\setcounter{equation}{0}
\section*{7 Conclusion}
Exact evolution equations provide a powerful tool for nonperturbative
calculations in quantum field theory.
Although it is not possible in practice to solve
them exactly, the method of truncating
the space of actions yields
nonperturbative answers which
require neither an expansion in the number of loops nor in any small
coupling constant. The approximation involved here is that
during the evolution the mixing of the operators retained in the
ansatz for $\Gamma_k$~ with all other operators is neglected.
The examples of QCD and of Chern-Simons theory which we discussed
in these notes illustrate that this approach works equally well for
theories with a complicated dynamics and for topological
theories.

\section*{Acknowledgement}
It is a pleasure to thank the organizers of the workshop for their
hospitality in Corfu and for the opportunity to present this
material there.

\end{document}